\documentclass[11pt]{article}
\usepackage{geometry}
\usepackage{color}
\usepackage{amsthm}
\usepackage{amsmath}
\usepackage{amsfonts}
\usepackage{graphicx}
\newtheorem{thm}{Theorem}[section]

\newtheorem{conj}[thm]{Conjecture}
\newtheorem{lem}[thm]{Lemma}

\newtheorem{claim}[thm]{Claim}
\theoremstyle{definition}
\newtheorem{defn}[thm]{Definition}

\theoremstyle{remark}

\numberwithin{equation}{section}

\newcommand{\eps}{\varepsilon}

\newcommand{\A}{\mathcal{A}}

\def\sign{\operatorname{sign}}

\def\ent12{\hat \Phi_{|1,2}}
\def\R{\mathbb R}
\def\eps{\varepsilon}

\renewcommand{\cos}{\operatorname{cos}}
\renewcommand{\sin}{\operatorname{sin}}
\title{The Complexity of Computing (Almost)  Orthogonal Matrices With $\eps$-Copies of the Fourier Transform\footnote{This work was supported by ERC grant SpeedInfTradeoff}}
\author{Nir Ailon \ \ \ Gal Yehuda\\ Department of Computer Science \\ Technion Israel Institute of Technology \\ Haifa, Israel }

\begin{document}

\maketitle

\def\sign{\operatorname{sgn}}
\def\dim{n}
\def\C{\mathbb C}
\def\R{\mathbb R}
\def\P{\mathcal P}
\def\trace{\operatorname{tr}}
\def\diag{\operatorname{diag}}
\def\rank{\operatorname{rank}}
\def\F{{\mathcal F}}
\def\Id{\operatorname{Id}}
\def\f{{\hat f}}
\def\A{{\mathcal A}}
\def\Z{{\mathbb Z}}

\begin{abstract}

The complexity of computing the Fourier transform is a longstanding open problem.  Very recently, Ailon (2013, 2014, 2015) showed in a collection of papers that, roughly speaking, a speedup of the Fourier transform computation  implies numerical ill-condition.  The papers also quantify this tradeoff. 
The main method for proving these results is via a potential function called quasi-entropy, reminiscent of Shannon entropy.    The quasi-entropy method opens new doors to understanding the computational complexity of the important  Fourier transformation.  However, it suffers from various obvious limitations.   
This paper is motivated by one such limitation, related to the computation of near-orthogonal matrices that have the Fourier transform `hidden' in low-order bits. While partly overcoming this limitation, the paper sheds light on new interesting, open problems
on the intersection of computational complexity and group theory.  The paper also explains why this research direction, if fruitful, has a chance of solving much bigger questions about the complexity of the Fourier transform.
\end{abstract}

\section{Introduction}\label{sec:intro}
The Fourier transform is one of the most important  linear transformations in science and engineering.
The (normalized) Discrete Fourier Transform (DFT) $\hat x\in\C^n$ for input signal $x\in \C^n$ is
defined by
$ \hat x_i = n^{-1/2}\sum_{j=1}^n e^{-2\pi \iota (i-1)(j-1)/n} x_j\ ,$
where $\iota = \sqrt{-1}$.   
 DFT has applications in many fields, including fast polynomial multiplication \cite[chapter~30]{Cormen:2009:IAT:1614191}, fast
integer multiplication \cite{Furer:2007:FIM:1250790.1250800}, fast large scale linear algebra and matrix sketching \cite{Avron:2010:BSL:1958627.1958633,sarlos2006improved},
signal processing \cite[chapters~ 6-9]{elliott2013handbook} and more.
From a theoretical perspective, the DFT is a special case  of the more general  Fourier transform
on  abelian groups,  with respect to the group  $\Z/n\Z$.   Another special case, known as the Walsh-Hadamard
transform, is defined over the group $(\Z/2\Z)^{\log_2 n}$ (for integer $\log_2 n$).
The Walsh-Hadamard transform $\hat x^{WH}$is given by 
$ \hat x^{WH}_i = n^{-1/2} \sum_{j=1}^n (-1)^{\langle i-1, j-1\rangle} x_j\ ,$
where $\langle a, b \rangle$ here is dot product of two bit vectors corresponding to the base-$2$ representation of
integers $a,b$.
The WH transform  has applications in coding theory and digital image processing  as well as  in fast large scale
linear algebra and matrix sketching \cite{boutsidis2013improved}.  It is also instrumental in analysis of boolean functions
\cite{blais2012testing,fischer2002testing} and more generally in theoretical computer science and learning theory \cite{mansour1994learning}.

From a computational point of view, an $O(n\log n)$ algorithm is known for both the DFT and the WH
transform.  For the DFT case, this was discovered by Cooley and Tukey in 1965 \cite{1965-cooley} in a seminal
paper.   For the WH case, the corresponding WH transform has been discovered in 1976 \cite{10.1109/TC.1976.1674569}.
Both algorithms run in a linear algebraic computational model (more on that in Section~\ref{compmodel}).

The complexity of computing the Fourier transform is, on the other hand a longstanding open problem, hence every contribution and insights are important. 
An $\Omega(n)$ bound is trivial due to the necessity to consider all input coordinates.
The gap between $\Omega(n)$ and $O(n\log n)$ may seem small.  Nevertheless, owing to the
importance of the Fourier transform (both DFT and WH), it is crucial to close it in a reasonable model of computation.  Some early
results \cite{Morgenstern:1973:NLB:321752.321761} proved a lower bound of the \emph{unnormalized} Fourier transform,
defined as a scaling up of the Fourier transform by $n^{1/2}$,
using a potential function that is related to matrix determinant.  
 This result, though shedding light on an important problem, unfortunately does not explain why the normalized (orthogonal)  Fourier transform has complexity $\Omega(n\log n)$ and, conversely, does not explain
why the Fourier transform is computationally more complex than a scaling of the identity matrix by a
factor $n^{1/2}$.
A result by Papadimitriou \cite{Papadimitriou:1979:OFF:322108.322118} provides a lower bound
of computing the Fourier transform in finite fields in a computational model that defines
a certain information flow graph.  The result does not seem to be applicable to the real case.

 Recently
Ailon \cite{DBLP:journals/cjtcs/Ailon13,DBLP:conf/icalp/Ailon15,DBLP:journals/toct/Ailon16} showed in a collection of papers that speedup of Fourier computation (both DFT and WH) implies
ill-conditioned computation (see also Section~\ref{compmodel} below for a precise definition).   The result uses a potential 
reminiscent of Shannon entropy function on probability vectors, except that it is applied to 
any real vector (including e.g. negative numbers), hence the `quasi' adornment.

\subsection{Computational Model and the Quasi-Entropy Method}\label{compmodel}
In this work we work over the reals $\R$, and when discussing the complex matrix DFT of order $n$, we think of its real embedding
of order $2n$.
An algorithm $\A$ computing a real $n$ by $n$ matrix $M$ in $m$ steps is a sequence $(\Id_n=M^{(0)}, M^{(1)}, \dots, M^{(m)}=M)$ of matrices, where for each $t\in [1,m]$, $M^{(t)}$ is obtained from $M^{(t-1})$ by one of two ways:
\begin{enumerate}
\item[(1)] Planar rotation matrix: $M^{(t)} = R^{(t)}M^{(t-1)}$, where  $R^{(t)}$ is defined using two indices $i = i^{(t)}$, $j=j^{(t)}$ and
an angle $\Theta = \Theta^{(t)}$ by
$\left ( \begin{matrix} R_{i,i}& R_{i,j} \\R_{j,i} & R_{j,j} \end{matrix}\right)=\left ( \begin{matrix} \cos\Theta & \sin \Theta \\ -\sin\Theta & \cos \Theta \end{matrix}\right)$, the remaining diagonal coordinates by $1$ and all other coordinates by $0$.
\item[(2)] Constant gate matrix: $M^{(t)} = C^{(t)}M^{(t-1})$, where $C^{(t)}$ is defined using an index $i=i^{(t)}$ and a nonzero constant
$c=c^{(t)}$ as a diagonal matrix with the $i$'th diagonal element equalling $c$ and the rest as $1$. 
\end{enumerate}
An algorithm runs in the \emph{orthogonal model} if it performs steps of type (1) only.\footnote{In previous work, this was called \emph{unitary model}, but we use the terminology orthogonal, which is standard and technically more accurate when working over the reals.}
Applying an algorithm $\A$ in this model to an input vector $x$, saved in a corresponding buffer of $n$ numbers, is done by iteratively performing at step $t$ either an in-place planar rotation w.r.t angle $\Theta$ on two buffer elements at positions $i^{(t)}, j^{(t)}$, or in-place multiplying a single buffer element at
position $i^{(t)}$ by $c=c^{(t)}$.  The resulting buffer content is $Mx$.

The condition number $\kappa(M)$ of a matrix $M$ is the ratio between the top and the bottom singular values.  It is $\kappa$-well conditioned if $\kappa(M)\leq \kappa$.    Otherwise it is $\kappa$-ill conditioned.  The condition number 
$\kappa(\A)$ of
an algorithm $\A$ is $\max_{t\in[1,m]} \kappa(M^{(t)})$.  It is  $\kappa$-well conditioned if  $\kappa(\A) \leq \kappa$.  Otherwise
if is $\kappa$-ill conditioned.

We briefly remind the reader of (preconditioned) matrix quasi-entropy:  For a nonsingular real $n$-by-$n$ matrix $M$, the matrix quasi-entropy $\Phi_{A,B}(M)$ with respect to two matrices $A,B$ of $n$ rows and a matching number $\nu$ of columns is given as
\begin{equation}\label{eq:defprecond}
\Phi_{A,B}(M) = -\sum_{i=1}^n \sum_{j=1}^\nu (MA)_{i,j} (M^{-T} B)_{i,j} \log |(MA)_{i,j} (M^{-T} B)_{i,j}|\ .
\end{equation}
where $M^{-T}$ is shorthand for inverse-transpose.  Throughout all logarithms are base $2$. 
If $A=B=\Id_n$ then we write $\Phi(M)$.
If additionally $M$ is orthogonal then $\Phi(M) = -\sum_{i,j} |M_{i,j}|^2\log|M_{i,j}|^2\ .$  In such a case, we simply say `entropy' instead of `quasi-entropy'.\footnote{The reason for the term \emph{quasi-entropy} is that the expressions $M_{i,j}M^{-T}_{i,j}$ may be
negative, or greater than $1$, and hence $\Phi$ is an extension of the usual entropy function, applied to probabilities.}
For an orthogonal matrix $M$ of order $n$, we say that $M$ has high entropy if $\Phi(M) = \Omega(n \log n)$.
\noindent
The main technical lemma  in \cite{DBLP:journals/toct/Ailon16} is:
\begin{lem}[From  \cite{DBLP:journals/toct/Ailon16}]\label{lemailon16B} 
If $M^{(t)}$ is obtained from $M^{(t-1)}$ by a planar rotation matrix and $\kappa = \kappa(M^{(t)})$, then
$ \left | \Phi_{A,B}(M^{(t)}) - \Phi_{A,B}(M^{(t-1)}) \right | \leq 2\kappa\cdot  \|A\|\cdot\|B\|$.
If $M^{(t)}$ is obtained from $M^{(t-1)}$ by a constant gate matrix, then $\Phi_{A,B}(M^{(t)}) = \Phi_{A,B}(M^{(t-1)})$.
\end{lem}

The quasi-entropy of the $n$-by-$n$ identity matrix  $\Id_n$ is $0$, that of the Fourier matrix
(both DFT and WH) is $\Omega(n \log n)$.  Therefore, in the uniformly $\kappa$-well conditioned
 a lower bound of
the number of steps is $\Omega(\kappa^{-1}n\log n)$. Equivalently, a speedup of FFT by factor of
$\kappa>1$ implies $\kappa$-ill conditioned computation.  %

\subsection{Our Contribution}\label{sec:problemwithmethod}

The aforementioned result, based on the quasi-entropy method, is limited. 
The result implies that if we sped up FFT by a factor of $\kappa$, then the resulting computation would be $\Omega(\kappa)$-ill conditioned.  This is a mild implication, and the author conjectures that the correct implication is a lower bound of $\Omega(\exp\{\kappa\})$ in the condition number requirement.  For an exact statement of the conjecture, we refer the reader to the aforementioned papers as
well as to a more recent  result \cite{paraunitary}.
Here we discuss and partially resolve a different problem, and also find an intimate connection between this problem and the
limitation just described.
Assume that instead of computing $Fx$, given an input vector $x$, we compute
$(\Id+\eps F) x\ ,$
where $\eps$ is a small constant.  
More precisely:
\begin{defn}
A matrix $M$ is   an  $\eps$-perturbation with respect to an orthogonal matrix $F$ 
 if $M = \Id + \eps F$.
\end{defn}
We will be interested in the case in which $F$ is a Fourier transform, or more generally, an orthogonal $C$-dense matrix.
Fix $\eps>0$ and let $M = \Id+\eps F$ be a corresponding $\eps$-perturbation with respect to $F$.
The matrix $M$ is not necessarily orthogonal, but it has condition number at most $1+O(\eps)$.  There also exists a $(1+O(\eps))$-well conditioned
algorithm computing $Mx$ given input vector $x$ (See Appendix~\ref{sec:can_compute} for proof).
Our main results are as follows:
\begin{claim}\label{theclaim}
The number of steps required for computing $M$ in a $(1+O(\eps))$-well conditioned algorithm is $\Omega(\eps n \log n)$.
\end{claim}
\begin{thm}\label{mainthm}
In case $F$ is symmetric (e.g. the Hadamard matrix),  the number of steps required for computing $M$  in a $(1+O(\eps))$-well conditioned algorithm is
$ \Omega\left ( \frac{ n \log n}{\log \eps^{-1}}\right )\ .$ 
\end{thm}
\noindent
Theorem~\ref{mainthm} also holds if $F$ is almost symmetric.\footnote{ Almost symmetric means that 
$ \frac {\left \|\frac{F + F^T}{2}\right\|_F^2 } { \|F\|_F^2} = \Omega(1)\ $.  This holds for  the (real representation of) DFT.  We do not elaborate on this simple extension here.}
Our main conjecture is as follows:
\begin{conj}\label{theconj}
The number of steps required for computing $M$  in a $(1+O(\eps))$-well conditioned algorithm is
$ \Omega\left ( \frac{ n \log n}{\log \eps^{-1}}\right )$
(\emph{unconditionally} on symmetry properties of $F$).
\end{conj}

\subsection{Justification for Studying this Problem}

Understanding the complexity of computing an $\eps$-perturbation of $F$ seems like a toy problem, but it is
important to study it.  We now explain why.

\subsubsection*{Reason 1: Free lunch by reducing $\eps$?!}
Assume that we could compute an $\eps$-perturbation $M$ with respect to a Fourier transform matrix $F$, with
some small parameter $\eps$, much faster than the time it would take to compute $F$.  The implications would be 
that, we could compute $y : = Mx = x + \eps Fx$ quickly, then compute $y-x$ (after, say, having stored $x$ somewhere)
and finally output $(y-x)/\eps = Fx$.  But of course it seems extremely unlikely that we could obtain a significant speedup 
simply by choosing a small $\eps$.  This is the main reason for Conjecture~\ref{theconj}.    The fact that the proof of
Theorem~\ref{mainthm} works only for a symmetric $F$ is quite odd.  

\subsubsection*{Reason 2: The bigger picture}

In \cite{paraunitary}, the author presents an \emph{algebraic reduction} from an algorithm computing $F$ in a $\kappa$-well conditioned
model, to another algorithm that runs in the orthogonal model, and computes a matrix $\tilde M = \tilde \Id + \frac {1}{\sqrt \kappa} \tilde F\ ,$ where $\tilde I$ is a sparse (low entropy) matrix, and $\tilde F$ is an orthogonal matrix of high entropy (that is related to $F$ in a complicated way that is
beyond the scope of this work).  If Conjecture~\ref{theconj} were true, then it would indicate that the number of steps
required to compute $\tilde M$ is $\Omega((n\log n)/\log \kappa)$.  This would be a major breakthrough, because it
would imply that speeding up FFT by a factor of $s$ requires condition number $\kappa = \exp\{\Omega(s)\}$, exponentially better than the current bound, with yet further heavy implications on the \emph{bit}-operation cost incurred by any such (theoretical) speedup.
Here  we present the ``toy'' problem, partly solve it, and keep the bigger picture in mind.

\subsection{Proof of Claim~\ref{theclaim}}
Assume  an algorithm $\A=(\Id=M^{(0)}, M^{(1)}, \dots, M^{(m)}=(\Id+\eps F))$ computes the Fourier $\eps$-perturbation, and that $\A$ is $(1+O(\eps))$-well conditioned.
We will choose a pair of preconditioning matrices $A,B$ that are orthogonal, for which
\begin{equation}\label{pppppppp}  |\Phi_{A,B}(\Id)| = o(\eps  n\log n)\ \ \ \ \ \  \Phi_{A,B}(\Id+\eps F) = \Omega(\eps n \log n).\end{equation}

Using Lemma~\ref{lemailon16B} (with the spectral norm  bound for  $A,B$) implies that $\forall t\in [m]$:
$$ \left | \Phi_{A,B}(M^{(t)}) - \Phi_{A,B}(M^{(t+1)}) \right | = O(1)\ .$$
Combining, the implication is that $m = \Omega(\eps n \log n)$.
It turns out that (\ref{pppppppp}) can be achieved by taking $A=\Id$ and $B=F+\eps \Id$. It is easy to check that
\begin{eqnarray}
  |\Phi_{A,B}(\Id)| &=& |\Phi_{\Id, F+\eps \Id }(\Id)|
= \left | \sum_{i=1}^n (F_{i,i}+\eps)  \log | F_{i,i}+\eps |  \right |
\leq 2n\ , \nonumber
\end{eqnarray}
where the rightmost inequality is because of the observation that $\max_{0\leq x\leq 1+\eps} |x|\log|x|$ is (loosely) at most $2$.
\begin{eqnarray}
& & \Phi_{\Id,F+\eps \Id}(\Id + \eps F) = - \sum_{i = 1}^{n} \sum_{j=1}^n L( (\Id + \eps F)_{i,j}\cdot ((\Id + \eps F)^{-T}(F+\eps \Id))_{i,j}) \nonumber  \\
& &\hspace{0.2cm} = - \sum_{i,j = 1}^{n} L( (\Id + \eps F)_{i,j}\cdot ((\Id + \eps F^T)^{-1}(\Id+\eps F^T)F)_{i,j})
  = - \sum_{i,j = 1}^{n}  L( (\Id + \eps F)_{i,j}\cdot F_{i,j}) \nonumber \\
& &\hspace{0.2cm}= -\sum_{i\neq j} \eps F_{i,j}^2\log |\eps F_{i,j}^2| - \sum_{i=1}^n (1+\eps F_{i,i})F_{i,i}\log|(1+\eps F_{i,i})F_{i,i} |\  \nonumber \\
& &\hspace{0.2cm}\geq  -\sum_{i\neq j} \eps F_{i,i}^2\log |\eps F_{i,i}^2| - 2n
= -\sum_{i,j} \eps F_{i,j}^2\log|\eps F_{i,j}^2| + \sum_i \eps F_{i,i}^2\log|\eps F_{i,i}^2|- 2n \nonumber \\
& &\hspace{0.2cm}\geq  \eps \Phi(F)  - 4n  = \Omega(\eps n\log n) \ , \nonumber
\end{eqnarray}
where the first and second inequalities are, again,  from the last observation.
This concludes the proof.
\subsection{Proof of Theorem~\ref{mainthm}}
As assumed in the theorem, $F$ is symmetric.
We will use a sightly different  preconditioned potential function.
For a nonsingular matrix $M$ and two $n$-by-$2n$ matrices $P,Q$, we define
\begin{eqnarray}
\hat \Phi_{P,Q}(M) &=& -\sum_{i=1}^n \sum_{j=1}^n L((MP)_{i,j}(M^{-T}Q)_{i,j} +(MP)_{i,j+n}(M^{-T}Q)_{i,j+n} ) 
\end{eqnarray}  

\noindent
We will use $\hat \Phi$ in conjunction with the preconditioners
$
P = [\Id,\,-F]\ ,
Q = [F,\,\Id] \ ,
$
where  for matrices $A,B$ with compatible number of rows, $[A, B]$ is the matrix obtained by stacking $B$ to the right of $A$.
We aim to prove that a rotation can change the potential by at most $O(\eps \log \eps^{-1})$.
(It is clear to see that
a constant gate does not change the potential.)
Let $t$ be such that $M^{(t+1)}$ is obtained from $M^{(t)}$ by left multiplication by a planar rotation matrix $R^{(t)}$
with respect to indices $i,i'$ and
angle $\Theta$.  Without loss of generality we can assume that $i=1, i'=2$.
First, one can notice that since $\kappa(M^{(t)}) = 1 + O(\eps)$, 
\begin{equation}
M^{(t)} = U+\Delta ,\ \ \label{eq:MUDelta} 
{(M^{(t)})}^{-T} = U + \Gamma 
\end{equation}
where $U$ is an orthogonal matrix, and $\Delta,\Gamma$ have spectral norm $O(\eps)$.
 We can ignore the contribution  to the potential coming from rows
$i>2$.
Let $V = UF, \Xi = \Delta F,  \Lambda = \Gamma F$.
We denote by $\ent12$  the contribution to $\hat \Phi_{P,Q}(M^{(t)})$ coming from rows $i=1,2$, namely: 
\begin{eqnarray}
\ent12 &=&\sum_{j=1}^n L((U_{1,j}+\Delta_{1,j})(V_{1,j} + \Lambda_{1,j}) - (V_{1,j}+\Xi_{1,j})(U_{1,j}+\Gamma_{1,j})) \nonumber\\
&-& 
\sum_{j=1}^n L((U_{2,j}+\Delta_{2,j})(V_{2,j} + \Lambda_{2,j}) - (V_{2,j}+\Xi_{2,j})(U_{2,j})+\Gamma_{2,j}))\ . \nonumber 
\end{eqnarray}
The term $U_{1,j}V_{1,j}$ disappears from the first row, and $U_{2,j}V_{2,j}$ from the second.
Hence, 
\begin{eqnarray}
\ent12&=&  -\sum_{j=1}^n L(  U_{1,j}\Lambda_{1,j} + V_{1,j}\Delta_{1,j} + \Delta_{1,j}\Lambda_{1,j} 
-    V_{1,j}\Gamma_{1,j} - U_{1,j}\Xi_{1,j}  - \Gamma_{1,j}\Xi_{1,j})  \nonumber  \\
&-& \sum_{j=1}^n L(  U_{2,j}\Lambda_{2,j} + V_{2,j}\Delta_{2,j} + \Delta_{2,j}\Lambda_{2,j} 
 -    V_{2,j}\Gamma_{2,j} - U_{2,j}\Xi_{2,j}  - \Gamma_{2,j}\Xi_{2,j})\ .\nonumber
\end{eqnarray}
\noindent
Define:
$$r_j = \sqrt{U_{1,j}^2 + V_{1,j}^2 + U_{2,j}^2 + V_{2,j}^2},\ \ \  
\rho_j = \sqrt{\Delta_{1,j}^2 + \Xi_{1,j}^2 + \Delta_{2,j}^2 + \Xi_{2,j}^2 +\Gamma_{1,j}^2 + \Lambda_{1,j}^2 + \Gamma_{2,j}^2 + \Lambda_{2,j}^2 }\ .$$ 

Note that by our construction we have:
$\sum_{j=1}^n r_j^2 = 4 $ and 
$\sum_{j=1}^n \rho_j^2 = O(\eps)$,
where the former  is by orthogonality of $U,V$ and the latter bound is by the spectral bound of $O(\eps)$
on $\Delta, \Gamma, \Delta, \Xi$ and $\Lambda$.
Dividing and multiplying by $r^2_j$, we get: 

\begin{eqnarray}
\ent12&=&-\sum_{j=1}^n L\left (r^2_j\left ( \frac{ U_{1,j}\Lambda_{1,j}}{r^2_j} + \frac{V_{1,j}\Delta_{1,j}}{r^2_{j}} + \frac{\Delta_{1,j}\Lambda_{1,j}}{r^2_j}  \label{eq:dividemult}
 -    \frac{V_{1,j}\Gamma_{1,j}}{r^2_j} - \frac {U_{1,j}\Xi_{1,j}}{r^2(j)}  - \frac{\Gamma_{1,j}\Xi_{1,j}}{r^2_j} \right )\right ) \nonumber \\
&-& 
\sum_{j=1}^n L\left ( r^2_j \left (\frac{U_{2,j}\Lambda_{2,j}}{r^2_j} + \frac{V_{2,j}\Delta_{2,j}}{r^2_{j}} + \frac{\Delta_{2,j}\Lambda_{2,j}}{r^2_j}
 -   \frac{ V_{2,j}\Gamma_{2,j}}{r^2_j} - \frac{U_{2,j}\Xi_{2,j}}{r^2_j}  - \frac{\Gamma_{2,j}\Xi_{2,j}}{r^2_j}\right )\right )\ . \nonumber
\end{eqnarray}

For simplicity of notation, for rows $i=1,2$  and any column $j$ we define:
\begin{eqnarray}
X_{i,j} &=&  \frac{ U_{i,j}\Lambda_{i,j}}{r^2_j} + \frac{V_{i,j}\Delta_{i,j}}{r^2_j} + \frac{\Delta_{i,j}\Lambda_{i,j}}{r^2_j} \label{eq:Xij}
 -    \frac{V_{i,j}\Gamma_{i,j}}{r^2_j} - \frac {U_{i,j}\Xi_{i,j}}{r^2_j}  - \frac{\Gamma_{i,j}\Xi_{i,j}}{r^2_j} \ ,
\end{eqnarray}
so that
$\ent12 =   -\sum_{j=1}^n L(r_j^2 X_{1,j})    -\sum_{j=1}^n L(r_j^2 X_{2,j})$.
We now write $M^{(t+1)} = U'+\Delta' $ and ${(M^{(t+1)})}^{-T} = U' + \Gamma'$,
where $U'=R^{(t)}U$, $\Delta' = R^{(t)}\Delta$, $\Gamma' = R^{(t)}\Gamma$.
Similarly, we define $V' = R^{(t)}V$, $\Xi' = R^{(t)}\Xi$, $\Lambda' = R^{(t)}\Lambda$
as the `post-rotation' version of the corresponding variables.  We can  also define $r'_j, \rho'_j$ 
similarly to $r_j$ and $\rho_j$, but with the  post-rotation variables.  However  clearly
$r_j = r'_j$ and $\rho_j=\rho'_j$ because rotation is an isometry, so 
$U_{1,j}^2+U_{2,j}^2 = {U'}_{1,j}^2+{U'}_{2,j}^2$ 
and similarly
for the other components. 
The ultimate goal is to compare the corresponding potentials $\ent12$ and $\ent12'$, where $\ent12'$
is defined like $\ent12$, but using the post-rotation variables.
Now consider the expression $X_{1,j} + X_{2,j}$ for fixed $j$.  This expression can be viewed as sum of (scaled) inner products.  For example, the first inner product is 
$\frac{\left \langle (U_{1,j},U_{2,j}),\ (\Lambda_{1,j},\Lambda_{2,j})\right \rangle}{r^2_j}\ .$
Hence, 
\begin{equation}\label{eq:equivXs} X_{1,j}+X_{2,j} = X'_{1,j}+X'_{2,k}\ ,\end{equation} where $X'_{i,j}$ is obtained as in (\ref{eq:Xij}), but using the post-rotation variables.  Indeed planar inner products are not affected by a planar rotation.  
Therefore, $|\ent12 - \ent12'|$ equals
\begin{eqnarray}
& & 
\left |\sum_{j=1}^n (L(r_j^2X_{1,j})+ L(r_j^2X_{2,j}))
- \sum_{j=1}^n L(r_{j}^2X'_{1,j})+ L(r_j^2X'_{2,j}) )\right |\nonumber \\
&=&\left  |\sum_{j=1}^n \left (L(r_j^2) (X_{1,j}+X_{2,j}-X'_{1,j}-X'_{2,j}) 
 + r_j^2 (L(X_{1,j}) + L(X_{2,j}) - L(X'_{1,j}) - L(X'_{2,j})) \right )\right | \nonumber \\
&=& \left |\sum_{j=1}^n r_j^2 (L(X_{1,j}) + L(X_{2,j}) - L(X'_{1,j}) - L(X'_{2,j}))\right | \nonumber \\
&\leq& \sum_{j=1}^n r_j^2\left |(L(X_{1,j}) + L(X_{2,j}) - L(X'_{1,j}) - L(X'_{2,j})\right | \label{eq:last} \ ,
\end{eqnarray}

where the first equality is application of the rule $\log(xy)=\log x + \log y$, the second is from (\ref{eq:equivXs}) and the inequality is application of the triangle inequality.
Clearly, $|U_{i,j}|, |V_{i,j}| \leq r_j$ and $|\Lambda_{i,j}|, |\Delta_{i,j}|, |\Gamma_{i,j}|, |\Xi_{i,j}| \leq \rho_j$ for $i=1,2$
and $j\in[n]$.  Therefore by definition of $X_{i,j}$,  $|X_{i,j}| \leq 6 \rho_j/r_j$.  A similar bound
holds for $X'_{i,j}$.  For fixed $j$, each summand of  (\ref{eq:last}) is  hence bounded above by
$ r_j^2  (2\max_{|x| \leq 6\frac{\rho_j}{r_j} } L(x) - 2\min_{|y| \leq 6\frac{\rho_j}{r_j} } L(y)  ) \ .$

Let $e$ be the natural logarithm basis.
 For nonnegative $a$ such that $a \leq e^{-1}$, 
$\max_{x:|x|\leq a} L(x) = -|a|\log|a|$ and $\min_{x:|x|\leq a} L(x) =|a|\log|a|$. For $a > e^{-1}$,   $\max_{|x|\leq a} L(x) \leq a(3+\log a)$ and $\min_{|x|\leq a} L(x) \geq -a(3+\log a)$.
We will now define the following subsets of column indices, indexed by an integer $h$:
\begin{align*}
J_h = \begin{cases} 
\left  \{j: 6\frac{\rho_j}{r_j} \leq \eps\right \}   & h=0 \\
\left  \{j : 2^{h-1}\eps < 6\frac{\rho_j}{r_j} \leq \min\{2^{h}\eps, e^{-1}\} \right  \} & 
1\leq h \leq \lceil -\log (\eps e)\rceil  \\
\left  \{j : \max\{e^{-1}, 2^{h-1}\eps\} < 6\frac{\rho_j}{r_j} \leq 2^{h}\eps \right  \} & 
 h >  \lceil -\log (\eps e)\rceil \\
\end{cases}\ .
\end{align*}
(Notice that the collection $\{J_h\}$ is a disjoint cover of $[1,n]$).
Splitting  the sum (\ref{eq:last}) and applying our analysis of the function $L(x)$, we get
\begin{eqnarray} 
|\ent12 - \ent12'| &\leq&  \sum_{h\geq 0}\sum_{j\in J_h} r_j^2 \left (2\max_{|x| \leq 6\frac{\rho_j}{r_j} } L(x) - 2\min_{|y| \leq 6\frac{\rho_j}{r_j} } L(y) \right ) \nonumber \\ 
&\leq & -4\sum_{j\in J_0}r_j^2\eps\log \eps - 4\sum_{h=1}^{\lceil -\log (\eps e)\rceil}\sum_{j\in J_h} r_j^2  2^h \eps\log (2^h\eps) \\
& & +4\sum_{h > \lceil -\log (\eps e)\rceil}\sum_{j\in J_h} r_j^2\, 2^h \eps (3+\log (2^h \eps))\ .\label{eq:last2}
\end{eqnarray}

Let us now bound $\sum_{j\in J_h} r_j^2$ for all $h$.  For $h=0$ we trivially have 
$\sum_{j\in J_0} r_j^2\leq 4$.
 As for $h\geq 1$,
from the definition of $J_h$, we get that for all $j\in J_h$: 
$r^2_j \leq \frac{36\rho^2_j}{2^{2h-2}\eps^2}\ .$
Therefore,
$\sum_{j\in J_h} r^2_j \leq \sum_{j=1}^n \frac{36\rho^2_j}{2^{2h-2}\eps^2} \leq \tilde C2^{-2h}\ ,$ 
where $\tilde C$ is a global constant.
This gives:
\begin{eqnarray} 
|\ent12 - \ent12'| &\leq& -4\eps\log \eps -4 \sum_{h=1}^{\lceil -\log (\eps e)\rceil} \tilde C\,2^{-2h}  2^h \eps\log  (2^h \eps)  \nonumber\\
& & + 4\sum_{h > \lceil -\log (\eps e)\rceil}\tilde C\,2^{-2h} 2^h \eps (3+\log (2^h \eps))
=  O(-\eps \log \eps)\ . \nonumber
\end{eqnarray}
Finally, we will show that $\hat{\Phi}_{P,Q}(\Id)=0$ and $\hat{\Phi}_{P,Q}(\Id+\eps F) = \Omega(\eps n \log n)$.
\begin{eqnarray}
\hat \Phi_{P,Q}(\Id) &=&-\sum_{i=1}^n \sum_{j=1}^n (P_{i,j}Q_{i,j} +P_{i,j+n}Q_{i,j+n} )  \log |P_{i,j}Q_{i,j} +P_{i,j+n}Q_{i,j+n}| \nonumber 
\end{eqnarray}
By the definition of $P$ and $Q$, it holds that $P_{i,j}Q_{i,j} = -P_{i,j+n}Q_{i,j+n}$ and hence $\hat{\Phi}_{P,Q}(\Id) = 0$. 
We now prove that  ${\hat \Phi_{P,Q}(\Id +\eps F)} = \Omega(\eps n \log (n))$ . Write  $(\Id + \eps F)^{-1} = \Id - \eps F + Z$ where we notice that $Z$ has spectral norm $O(\eps^2)$.  Therefore $\hat \Phi_{P,Q}(\Id + \eps F) $ equals:
\begin{eqnarray}
& & -\sum_{i=1}^n \sum_{j=1}^n L((\Id+\eps F)_{i,j}(F-\eps F^2 +Z^TF)_{i,j} +(-(F+\eps F^2))_{i,j}(\Id - \eps F +Z^T)_{i,j} ) \nonumber  
\end{eqnarray}
The diagonal terms $(i=j)$ contribute $O(n)$ in absolute value, because the argument of $L(\cdot)$ for those terms is bounded in absolute value.
Hence, accounting for the off-diagonal:
\begin{eqnarray} 
\hat \Phi_{P,Q}(\Id + \eps F) &\geq& \underbrace {- \sum_{i, j}  L\left (\underbrace{2\eps F_{i,j}^2}_{\gamma_{i,j}}  + \underbrace{\eps F_{i,j} (Z^T F)_{i,j} - F_{i,j}Z^T_{i,j} - \eps (F^2)_{i,j}(Z^T_{i,j})}_{\delta_{i,j}} \right)}_{\Gamma} - O(n)  \nonumber  \ .%
\end{eqnarray}
(Technically the last summation should be only over the off-diagonal $(i\neq j)$, however it is cleaner to do the analysis over all pairs $i,j$.  The
difference can be `swallowed' by the $O(n)$ term.)
We estimate $\Gamma$ row by row.  
 Fix $i\in [n]$, and note 
that $\sum_j \gamma_{i,j} = 2\eps$ while $\sum_{j}| \delta_{i,j}| \leq 3\eps^2$, by the spectral bound on $Z$.
Therefore, intuitively, the $\gamma$ variables should dominate the sum, and by this intuition the first order
approximation would be $-\sum_{j} L(2\eps F_{i,j}^2)$.  This intuition is correct, and we now present the details.
Let $J_i$ be the set of indices $j\in[n]$ for which $|\delta_{i,j}|  \leq \gamma_{i,j} / 2$, and therefore $\gamma_{i,j} + \delta_{i,j} \geq{\gamma_{i,j}}/2$.  The function $-L(x)$ is monotonically increasing in the range $[-e^{-1}, e^{-1}]$ and therefore 
\begin{equation}\label{pqpqpq}
\forall j \in J_i:\ \ -L(\gamma_{i,j} + \delta_{i,j}) \geq -L(\gamma_{i,j}/2) = -L(\eps F_{i,j}^2)\ .
\end{equation} (We also used the fact that $|\gamma_{i,j}| + |\delta_{i,j}| \leq 5\eps$ which is
safely in the range $[-e^{-1}, e^{-1}]$ for small enough $\epsilon$.) Let $\bar J_i =[n] \setminus J_i$.  Then $\sum_{j\in \bar J_{i}}\gamma_{i,j} \leq 6 \eps^2$, because otherwise we would have $\sum_{j\in \bar J_{i}} |\delta_{i,j}| \geq \sum_{j\in \bar J_{i}} |\gamma_{i,j}|/2 > 6\eps^2 /2 = 3\eps^2$, a contradiction.  This gives the following estimate for $\Gamma$:
\begin{eqnarray}
\Gamma  &=& -\sum_{i,j} L(\gamma_{i,j} + \delta_{i,j}) = -\sum_{i}\sum_{j\in J_i} L(\gamma_{i,j} + \delta_{i,j})  -\sum_{i}\sum_{j\in \bar{J}_i} L(\gamma_{i,j} + \delta_{i,j}) \nonumber \\
&\geq& -\sum_{i}\sum_{j\in J_i} L(\gamma_{i,j}/2)  -\sum_{i}\sum_{j\in \bar{J}_i} L(\gamma_{i,j} + \delta_{i,j}) \nonumber \\
&=& -\sum_{i,j} L(\gamma_{i,j}/2) + \sum_i\sum_{j\in \bar J_i} L(\gamma_{i,j}/2)  -\sum_{i}\sum_{j\in \bar{J}_i} L(\gamma_{i,j} + \delta_{i,j}) \nonumber \\
&=&  -\sum_{i,j} L(\eps F_{i,j}^2) + \sum_i\sum_{j\in \bar J_i} L(\gamma_{i,j}/2)  -\sum_{i}\sum_{j\in \bar{J}_i} L(\gamma_{i,j} + \delta_{i,j}) \nonumber \\
&\geq& \eps \Phi(F) - \eps n\log \eps^{-1} - 20 \eps^2 n \log n = \Omega(\eps n \log n)\ ,
\end{eqnarray}
where we used (\ref{pqpqpq}) in the first inequality, and a loose estimation of $\sum_{j_\in \bar J_i} \gamma_{i,j}/2$ and  $\sum_{j_\in \bar J_i} |\gamma_{i,j} + \delta_{i,j}|$ as $10\eps^2 $ (each) in the last inequality.
(Recall also that we assumed $1/\eps$ is $n^{o(1)}$, hence $\log(1/\eps) = o(\log n)$.)
Combining this with the bound  $O(-\eps\log \eps)$ on the change of $\hat \Phi$ at each step
concludes the proof of the theorem.

\section{The Skew-Symmetric Case: An  Interesting Open Problem}
Assume now that $M$ is an $\eps$-perturbation of a skew-symmetric high-entropy orthogonal matrix $F$.
Then $M$ is an orthogonal matrix scaled up by $\sqrt{1+\eps^2}$ (because the eigenvalues of $F$ are all $\pm\iota$).
Hence it is mathematically more elegant to consider computation of $M^{(orth)} := M / \sqrt{1+\eps^2}$ instead.  
Indeed, it is natural to consider the complexity of $M^{(orth)}$ in the orthogonal model of computation, in which we
allow planar rotations only (without constant gate matrices).  This casts the problem as that of computing the distance between
two group elements $(\Id, M^{(orth)})$, with respect to a set of generators (planar rotations).
It is tempting to conjecture a lower bound of 
$ \Omega \left (\frac{n\log n}{\log (1/\eps)}\right )\ ,
$
as we achieved for the symmetric case,
but the proof technique developed for the symmetric case fails here.  
 In a nutshell, the reason for the failure is because the inverse-transpose of $M^{(orth)}$ is exactly $M$, 
while in the symmetric case the difference between $M$ and $M^{-T}$ is a \emph{nonzero} matrix of spectral norm $O(\eps)$.
The main open problem presented in this work is to improve the bound of $\Omega(\eps n\log n)$ for the skew-symmetric case.

\bibliographystyle{plain}

\bibliography{low_bound_fft}

\begin{thebibliography}{10}

\bibitem{DBLP:journals/cjtcs/Ailon13}
Nir Ailon.
\newblock A lower bound for fourier transform computation in a linear model
  over 2x2 unitary gates using matrix entropy.
\newblock {\em Chicago J. Theor. Comput. Sci.}, 2013, 2013.

\bibitem{DBLP:conf/icalp/Ailon15}
Nir Ailon.
\newblock Tighter {F}ourier transform lower bounds.
\newblock In {\em Automata, Languages, and Programming - 42nd International
  Colloquium, {ICALP} 2015, Kyoto, Japan, July 6-10, 2015, Proceedings, Part
  {I}}, pages 14--25, 2015.

\bibitem{DBLP:journals/toct/Ailon16}
Nir Ailon.
\newblock An omega((n log n)/r) lower bound for {F}ourier transform computation
  in the r-well conditioned model.
\newblock {\em {TOCT}}, 8(1):4, 2016.

\bibitem{paraunitary}
Nir Ailon.
\newblock Paraunitary matrices, entropy, algebraic condition number and fourier
  computation.
\newblock {\em arXiv preprint arXiv:1609.03278}, 2016.

\bibitem{Avron:2010:BSL:1958627.1958633}
Haim Avron, Petar Maymounkov, and Sivan Toledo.
\newblock Blendenpik: Supercharging {L}{A}{P}{A}{C}{K}'s least-squares solver.
\newblock {\em SIAM J. Sci. Comput.}, 32(3):1217--1236, April 2010.

\bibitem{blais2012testing}
Eric Blais.
\newblock {\em Testing properties of Boolean functions}.
\newblock PhD thesis, US Army, 2012.

\bibitem{boutsidis2013improved}
Christos Boutsidis and Alex Gittens.
\newblock Improved matrix algorithms via the subsampled randomized {H}adamard
  transform.
\newblock {\em SIAM Journal on Matrix Analysis and Applications},
  34(3):1301--1340, 2013.

\bibitem{1965-cooley}
James~W. Cooley and John~W. Tukey.
\newblock {An algorithm for the machine calculation of complex {F}ourier
  series}.
\newblock {\em Mathematics of Computation}, 19:297--301, 1965.

\bibitem{Cormen:2009:IAT:1614191}
Thomas~H. Cormen, Charles~E. Leiserson, Ronald~L. Rivest, and Clifford Stein.
\newblock {\em Introduction to Algorithms, Third Edition}.
\newblock The MIT Press, 3rd edition, 2009.

\bibitem{elliott2013handbook}
Douglas~F Elliott.
\newblock {\em Handbook of digital signal processing: engineering
  applications}.
\newblock Academic press, 2013.

\bibitem{10.1109/TC.1976.1674569}
B.J. Fino and V.R. Algazi.
\newblock Unified matrix treatment of the fast {W}alsh-{H}adamard transform.
\newblock {\em IEEE Transactions on Computers}, 25(11):1142--1146, 1976.

\bibitem{fischer2002testing}
Eldar Fischer, Guy Kindler, Dana Ron, Shmuel Safra, and Alex Samorodnitsky.
\newblock Testing juntas [combinatorial property testing].
\newblock In {\em Foundations of Computer Science, 2002. Proceedings. The 43rd
  Annual IEEE Symposium on}, pages 103--112. IEEE, 2002.

\bibitem{Furer:2007:FIM:1250790.1250800}
Martin F\"{u}rer.
\newblock Faster integer multiplication.
\newblock In {\em Proceedings of the Thirty-ninth Annual ACM Symposium on
  Theory of Computing}, STOC '07, pages 57--66, New York, NY, USA, 2007. ACM.

\bibitem{GolubV2013}
G.~H. Golub and C.~F. van Loan.
\newblock {\em Matrix Computations}.
\newblock Johns Hopkins University Press, Baltimore, 4th edition, 2013.

\bibitem{mansour1994learning}
Yishay Mansour.
\newblock Learning boolean functions via the {F}ourier transform.
\newblock In {\em Theoretical advances in neural computation and learning},
  pages 391--424. Springer, 1994.

\bibitem{Morgenstern:1973:NLB:321752.321761}
Jacques Morgenstern.
\newblock Note on a lower bound on the linear complexity of the fast {F}ourier
  transform.
\newblock {\em J. ACM}, 20(2):305--306, April 1973.

\bibitem{Papadimitriou:1979:OFF:322108.322118}
Christos~H. Papadimitriou.
\newblock Optimality of the fast {F}ourier transform.
\newblock {\em J. ACM}, 26(1):95--102, January 1979.

\bibitem{sarlos2006improved}
Tamas Sarlos.
\newblock Improved approximation algorithms for large matrices via random
  projections.
\newblock In {\em Foundations of Computer Science, 2006. FOCS'06. 47th Annual
  IEEE Symposium on}, pages 143--152. IEEE, 2006.

\end{thebibliography}

\appendix

\section{Fourier $\eps$-Perturbation Can be Computed by a $(1+O(\eps))$-Well Conditioned Algorithm}\label{sec:can_compute}

For completeness, we prove the simple fact stated in the section title.  By the SVD theorem, the matrix $(\Id+\eps F)$ can be written as a product of three matrices $U\Sigma V^T$, where $U$ and $V$ are real orthogonal and $\Sigma$ is diagonal nonnegative, with the elements on the diagonal in the range $[1-\eps, 1+\eps]$.
Therefore, to compute $(\Id+\eps F)x$ we can first compute $V^T x$, using the well known fact that any real orthogonal matrix is a composition of $O(n^2)$  rotations (see Chapter 5 on \emph{Givens}  rotations in \cite{GolubV2013}).  
Continuing from there, we can compute $\Sigma V^T x$ using constant gates, one per coordinate. 
Finally, we get $U\Sigma V^T x$ be decomposing $U$ as $O(n^2)$ rotations.
Clearly this computation is $(1+O(\eps))$-well conditioned.

\end{document}